\documentclass[a4paper,11pt]{article}
\usepackage{pos}
\usepackage{subcaption}

\newcommand{\SU}{\mathrm{SU}}
\newcommand{\nstep}{n_{\mathrm{step}}}

\newcommand{\DKL}{D_{\mathrm{KL}}}
\newcommand{\ESS}{\mathrm{ESS}}

\newcommand{\dd}{\mathrm{d}}
\newcommand{\Pf}{\mathcal{P}_{\mathrm{f}}}
\newcommand{\Pre}{\mathcal{P}_{\mathrm{r}}}

\title{Sampling $\SU(3)$ pure gauge theory with Stochastic Normalizing Flows}

\author[a]{Andrea Bulgarelli}
\author[a]{Elia Cellini}
\author*[a]{Alessandro Nada}

\affiliation[a]{Dipartimento di Fisica,  Universit\'a degli Studi di Torino and INFN, Sezione di Torino, \\
  Via Pietro Giuria 1, I-10125 Turin, Italy}

\emailAdd{alessandro.nada@unito.it}

\abstract{Non-equilibrium Monte Carlo simulations based on Jarzynski's equality are a well-understood method to compute differences in free energy and also to sample from a target probability distribution without the need to thermalize the system under study. 
In each evolution, the system starts from a given base distribution at equilibrium and it is gradually driven out-of-equilibrium while evolving towards the target parameters. 
If the target distribution suffers from long autocorrelation times, this approach represents a promising candidate to mitigate critical slowing down. Out-of-equilibrium evolutions are conceptually similar to Normalizing Flows and they can be combined into a recently-developed architecture called Stochastic Normalizing Flows (SNFs).
In this contribution we first focus on the promising scaling with the volume guaranteed by the purely stochastic approach in the $\SU(3)$ lattice gauge theory in 4 dimensions; then, we define an SNF by introducing gauge-equivariant layers between the out-of-equilibrium Monte Carlo updates, and we analyse the improvement obtained as well as the inherited scaling with the volume. 
Finally, we discuss how this approach can be systematically improved and how simulations of lattice gauge theories in four dimensions for large volumes and close to criticality can be realistically achieved.}

\FullConference{The 41st International Symposium on Lattice Field Theory (LATTICE2024)\\
 28 July - 3 August 2024\\
Liverpool, UK\\}

\begin{document}
\maketitle

\section{Introduction}

Flow-based architectures provide an innovative approach to sample from a probability distribution for which standard local Monte Carlo algorithms are highly inefficient, e.g. they suffer from critical slowing down. The idea behind these architectures is straightforward and consists in building a transformation that maps a prior distribution (which is easy to sample from) to the desired target distribution. While this approach goes back to the concept of trivializing maps~\cite{Luscher:2009eq}, recent developments in the deep learning community provided a powerful set of tools to build these mappings in the form of Normalizing Flows (NFs)~\cite{rezende2015variational}, yielding impressive results in low-dimensional lattice field theories (see ref.~\cite{Cranmer:2023xbe} for a review).

A particularly challenging form of critical slowing down, for example in the case of the $\SU(3)$ lattice gauge theory in four dimensions, is the so-called topological freezing~\cite{Schaefer:2010hu}: when approaching the continuum limit, the autocorrelation times of topological observables increase exponentially with the inverse of the lattice spacing. 
Normalizing Flows appear to be a perfect candidate to mitigate this issue: in a flow-based approach, for example, one could map the theory at coarser lattice spacing (corresponding to an inverse bare coupling $\beta_0$) to a finer lattice spacing (with $\beta>\beta_0$).
However, the performances of NFs appear to scale poorly when the number of degrees of freedom (and/or the dimensionality of the model) increases~\cite{Abbott:2022zsh, Albandea:2023wgd}, making their application to realistic volumes in four dimensions a challenge.

Non-equilibrium Markov Chain Monte Carlo (NE-MCMC) simulations based on Jarzynski's equality, which have seen widespread applications in several contexts in lattice field theory in the last years~\cite{Caselle:2016wsw, Caselle:2018kap, Francesconi:2020fgi, Bulgarelli:2023ofi}, also provide a way to mitigate critical slowing down. In this case, the transformation from the prior to the target distribution happens through a non-equilibrium process, which is implemented by updating the system with a specific protocol and driving it out of equilibrium in a controlled way. Indeed, in ref.~\cite{Bonanno:2024udh} a severe topological freezing was mitigated by sampling the target distribution with this approach and a scaling of the length of these evolutions with the relevant degrees of freedom was found. 
Interestingly, these non-equilibrium simulations share a very similar framework with Normalizing Flows, and in refs.~\cite{Caselle:2022acb} they were naturally combined into a new architecture called Stochastic Normalizing Flows (SNFs)~\cite{wu2020stochastic, Caselle:2022acb}, that has been recently used for numerical simulations of effective string theory~\cite{Caselle:2024ent}. In particular, the updates of (NE-MCMC) are interleaved with the building blocks of Normalizing Flows and in this way a hybrid architecture that mixes deterministic and stochastic transformations is created.

In this contribution we introduce for the first time a SNF for the $\SU(3)$ Yang-Mills theory in four spacetime dimensions and we test it on a mapping from coarser to finer lattice spacings. We analyse the features characterizing the training of the parameters of this architecture and we compare its performances with NE-MCMC. Most importantly, we confirm the findings of ref.~\cite{Bonanno:2024udh} concerning the scaling of NE-MCMC and we find that the same applies to SNFs as well.

\section{Out-of-equilibrium evolutions}

In a non-equilibrium Markov Chain Monte Carlo (NE-MCMC) a sequence of configurations is generated using a transition probability $P_{c(n)}$ that changes along the transformation, i.e.
\begin{equation}
\label{eq:nemcmc_seq}
  U_0 \stackrel{P_{c(1)}}{\longrightarrow} \; U_1 \;
  \stackrel{P_{c(2)}}{\longrightarrow} \; U_2 \;
  \stackrel{P_{c(3)}}{\longrightarrow} \; \dots \;
  \stackrel{P_{c(\nstep)}}{\longrightarrow} \; U_{\nstep} 
\end{equation}

The first configuration $U_0$ is sampled from the prior distribution $q_0 = \exp(-S_0)/Z_0$, while the last transition probability $P_{c(\nstep)}$ is the one corresponding to the target distribution $p=\exp(-S)/Z$.
The transition probabilities $P_{c(n)}$ are defined by a set of changing parameters $c(n)$, called the \textit{protocol}, that controls the intermediate actions used to update the system.
Formally, the overall out-of-equilibrium evolutions is defined by a forward transition probability
\begin{equation}
    \Pf[U_0,\dots,U] = \prod_{n=1}^{\nstep} P_{c(n)} (U_{n-1} \to U_n).
\end{equation}

Using Crooks' theorem~\cite{Crooks_1999} one can re-express the ratio of $\Pf$ with the transition probability of the reverse transformation $\Pre$
\begin{equation}
\label{eq:crooks}
    \frac{q_0 \Pf[U_0,\dots,U]}{p \Pre [U,\dots,U_0]} = \exp (W - \Delta F)
\end{equation}
with $\Delta F = -\log Z/Z_0$ and $W$ being the (dimensionless) work performed on the system in a given transformation while changing the parameters controlling the protocol, i.e.
 \begin{equation}
 \label{eq:work}
     W = \sum_{n=0}^{\nstep-1} \left\{ S_{c(n+1)}\left[U_n\right] - S_{c(n)}\left[U_n\right] \right\}.
 \end{equation}
Integrating eq.~\eqref{eq:crooks} over all possible transformation yields Jarzynski's equality~\cite{Jarzynski1997}
 \begin{equation}
 \label{eq:jarzynski}
     \langle \exp \left( -W \right) \rangle_{\mathrm{f}} = \frac{Z}{Z_0} = \exp \left( -\Delta F \right).
 \end{equation}
The average $\langle \dots \rangle_{\mathrm{f}}$ is formally defined as
 \begin{equation}
     \langle \mathcal{A} \rangle_{\mathrm{f}} = \int \dd U_0 \dots \dd U \; q_0(U_0) \, \Pf[U_0,\dots, U] \; \mathcal{A}[U_0, \dots, U].
 \end{equation}

While Jarzynski's equality provides a powerful tool to compute directly (ratios of) partition functions, the scope of non-equilibrium MCMC is much wider. Indeed, it is possible to sample any vacuum expectation value on the target distribution $p$ with a reweighting-like formula:
\begin{equation}
\label{eq:rw}
    \langle \mathcal{O} \rangle_p = \frac{\langle \mathcal{O} \exp(-W) \rangle_{\mathrm{f}} }{\langle \exp(-W) \rangle_{\mathrm{f}} };
\end{equation}
on the right-hand side the work $W$ is calculated along the whole non-equilibrium evolution while the observable $\mathcal{O}$ is computed only on the last configuration $U_{\nstep}$. 

It is clear that any hope of tackling critical slowing down on the target distribution $p$ requires a careful evaluation of the effectiveness of NE-MCMC in sampling $p$: in particular, for relatively small values of $\nstep$ we expect an overlap problem to appear between $p$ and the generated distribution at the end of the non-equilibrium transformation, i.e. 
\begin{equation}
 q(U) = \int \dd U_0 \dots \dd U_{\nstep -1} \dots q_0 (U_0) \Pf[U_0,\dots, U],
\end{equation}
which is generally intractable.
One possible figure of merit is the Kullback-Leibler (KL) divergence between the forward and reverse transition probabilities, that we can write as
\begin{equation}
\label{eq:kl}
 \DKL(q_0 \Pf \| p \mathcal{P}_{\mathrm{r}}) = \langle W \rangle_{\mathrm{f}} + \log \frac{Z}{Z_0}  = \langle W \rangle_{\mathrm{f}} - \Delta F.
\end{equation}
This quantity measures the similarity between the same evolution running forwards and backwards for a given protocol $c(n)$: as expected, if it is zero, then the transformation is reversible and $\langle W \rangle_{\mathrm{f}} = \Delta F$\footnote{The positivity of the KL divergence implies that $\langle W \rangle_{\mathrm{f}} \geq \Delta F$, i.e. the Second Law of thermodynamics for these types of MCMC transformations out of equilibrium}. Furthermore, it is easy to derive that $\DKL (q \| p) \leq \DKL(q_0 \mathcal{P}_{\mathrm{f}} \| p \mathcal{P}_{\mathrm{r}})$, thus putting an upper bound on the similarity of $q(U)$ and $p(U)$.

Another relevant metric we use to evaluate out-of-equilibrium evolutions is the Effective Sample Size, defined in this context as
\begin{equation}
     \hat\ESS = \frac{\langle \exp (-W) \rangle_{\mathrm{f}}^2}{\langle \exp(-2W) \rangle_{\mathrm{f}}}.
\end{equation}
It provides an approximation of the ratio of the variance of an observable $\mathcal{O}$ computed with the estimator of eq.~\eqref{eq:rw} on independent evolutions and the variance of the same observable computed on independent configurations sampled directly from $p$.

\section{Stochastic Normalizing Flows for $\SU(3)$ Yang-Mills theory}

Stochastic Normalizing Flows~\cite{wu2020stochastic, Caselle:2022acb} are an extremely promising architecture that can systematically improve the efficiency of out-of-equilibrium evolutions. In particular, in each evolution non-equilibrium MCMC updates are alternated with so-called \textit{coupling layers} (here denoted as $g_l$), i.e. deterministic transformations that are generally used to build Normalizing Flow architectures.
Compared to a NE-MCMC (see eq.~\eqref{eq:nemcmc_seq}), a sequence of configurations in a forward pass in a SNF looks like the following
\begin{equation}
   U_0 \stackrel{g_1}{\longrightarrow} \; g_1(U_0) \;
  \stackrel{P_{c(1)}}{\longrightarrow} \; U_1 \;
  \stackrel{g_2}{\longrightarrow} \; g_2(U_1) \;
  \stackrel{P_{c(2)}}{\longrightarrow} \; U_2 \;
  \stackrel{g_3}{\longrightarrow} \; \dots \;
  \stackrel{P_{c(\nstep)}}{\longrightarrow} \; U_{\nstep}
\end{equation}
where in each step a configuration is first transformed using the layer $g_n$ and then updated using a Monte Carlo algorithm with transition probability $P_{c(n)}$.

In this work we use gauge-equivariant coupling layers $g_l$ inspired by the ones introduced in refs.~\cite{Tomiya:2021ywc, Abbott:2023thq}. 
Each link is transformed according to a stout-smearing transformation~\cite{Morningstar:2003gk}, applied using 8 masks (4 for each space-time direction and 2 for even-odd sites) to ensure invertibility and a manageable computation of the Jacobian.
In particular we have that each active link $U_\mu (x)$ is transformed as
\begin{equation}
    U'_\mu (x) = g_l (U_\mu (x)) = \exp \left( Q_\mu^{(l)} (x) \right) \, U_\mu (x)
\end{equation}
with
\begin{equation}
    Q_\mu^{(l)} (x) = 2 \left[ \Omega^{(l)}_\mu (x)\right]_{\mathrm{TA}}
\end{equation}
where $\mathrm{TA}$ is the traceless and antihermitian operation. 
$\Omega_\mu^{(l)} (x)$ is a sum of untraced plaquettes, namely
\begin{equation}
    \Omega_\mu^{(l)} (x) = C_\mu^{(l)} (x) U^\dagger_\mu (x).
\end{equation}
Here $U_\mu(x)$ is the active link, while the $C_\mu^{(l)} (x)$ is a weighted sum of the staples $S_{\mu\nu}(x)$, that are composed exclusively of frozen links, i.e. links which are not transformed in the current mask:
\begin{equation}
\label{eq:cmu_stout}
    C_\mu (x) = \sum_{\nu \neq \mu} \rho^{(l)}_{\mu \nu}(x) S_{\mu\nu} (x)
\end{equation}

The stout-smearing parameters $\rho^{(l)}_{\mu \nu}(x)$ in eq.~\eqref{eq:cmu_stout} have to be determined using a training procedure. In this work we simply set them to be invariant under translations and (discrete) rotations, i.e. $\rho^{(l)}_{\mu \nu}(x)=\rho^{(l)}$; we also set them to be shared by all masks in each layer $l$.
Concerning the training procedure itself, we minimize the same KL divergence of eq.~\eqref{eq:kl}, with the only difference being the use of a "generalized" work, defined as
\begin{equation}
\label{eq:work_snf}
    W = \sum_{n=0}^{\nstep-1} S_{c(n+1)} (g_n(U_n)) - S_{c(n)} (g_n(U_n)) - \log \left| \det J_n (U_n) \right|
\end{equation}
which takes into account the logarithm of the determinant of the Jacobian $J_n$ of the coupling layer $g_n$.
Using this definition of work, the formulae of the previous section still hold: indeed SNFs can be used both for the computation of ratios of partition functions using Jarzynski's equality~\eqref{eq:jarzynski} and for the sampling of generic observables using eq.~\eqref{eq:rw}.

\section{Numerical results}

We test the scaling of both NE-MCMC and SNFs on the $\SU(3)$ pure lattice gauge theory in 4 dimensions for transformations in the lattice spacing. We use the plaquette (Wilson) action and the Monte Carlo update of choice during the out-of-equilibrium evolutions is composed of 1 step of heatbath plus 4 steps of over-relaxation.
The system is thermalized at a given initial value of the inverse bare coupling $\beta_0$ and then it is driven out of equilibrium using a linear protocol in $\beta$: 
\begin{equation}
  \beta(n) = \beta_0 + n \; \frac{\beta_t - \beta_0}{\nstep},
\end{equation}
with $\beta_t>\beta_0$ being the target value of the inverse bare coupling.

We studied lattice sizes $L/a=[10,12,16,20]$ and we focused our effort on a protocol in $\beta$ from $\beta_0=6.02$ to $\beta_t=6.178$. This approximately corresponds to a change in the physical size of the system of $1.8 \mathrm{fm} \to 1.4 \mathrm{fm}$ for a $L/a=20$ lattice. We remark that perfectly compatible results have been found also for the $\beta=5.896 \to 6.037$ protocol\footnote{We remark that at these values of the inverse coupling, using a heatbath+overrelaxation algorithm on a standard equilibrium simulation, topological observables are not frozen yet. We leave the application of these architectures on finer lattice spacings (and correspondingly larger volumes) to future work.}.

\begin{figure}[h]
 \centering
 \begin{subfigure}{.5\textwidth}
 \includegraphics[scale=0.5,keepaspectratio=true]{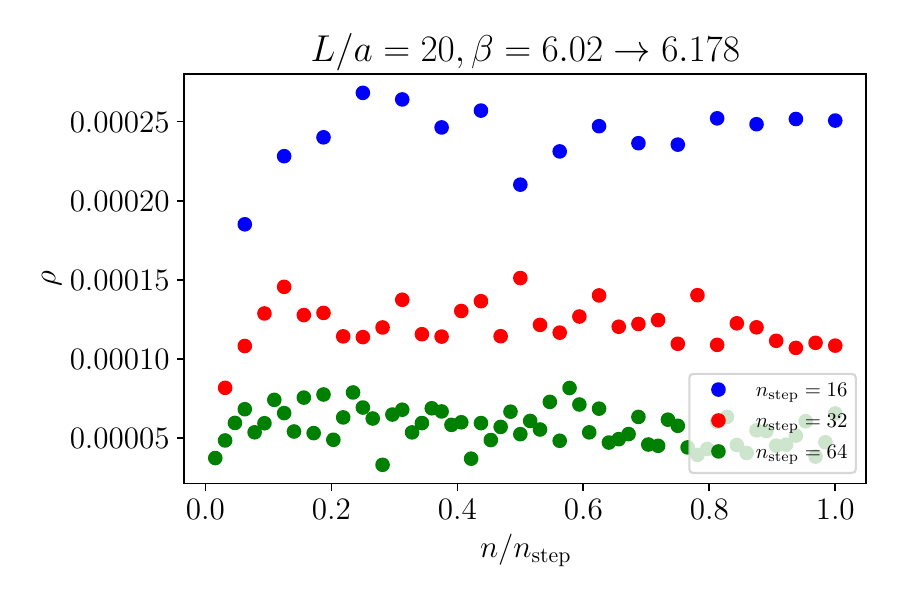}
 \end{subfigure}%
 \begin{subfigure}{.5\textwidth}
 \includegraphics[scale=0.5,keepaspectratio=true]{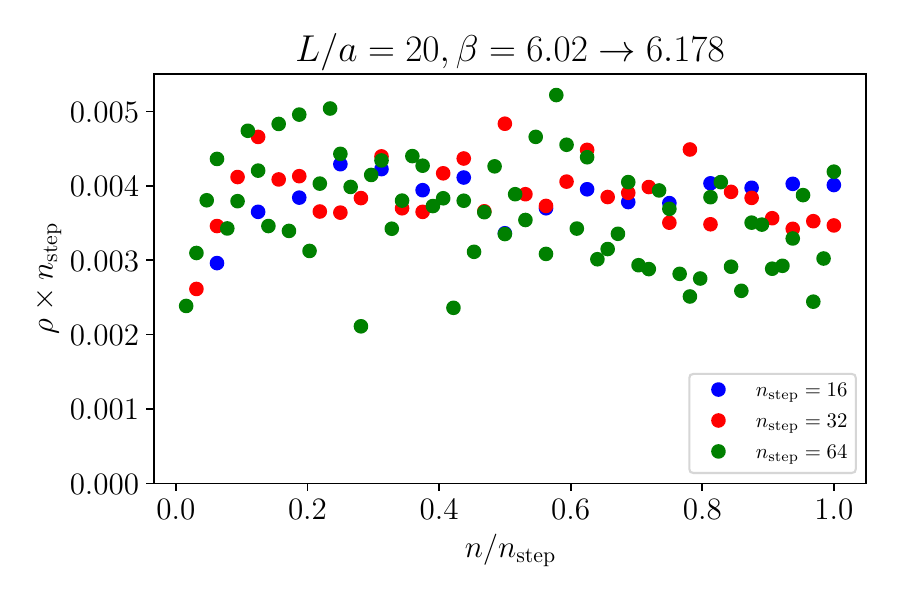}
 \end{subfigure}
 \caption{Value of the learned parameter $\rho$ along the SNF for the $n$-th layer (left panel) and the same value rescaled for $\nstep$ (right panel), for a training performed on a $L/a=20$ lattice for three values of $\nstep$.}
 \label{fig:rho}
\end{figure}

In order to find the optimal parameters of the SNF we minimize the KL divergence of eq.~\eqref{eq:kl} using an Adam optimizer. The peculiar structure of the work of eq.~\eqref{eq:work_snf} is such that we can train along each forward pass each layer separately, in a procedure reminiscent of ref.~\cite{Matthews:2022sds}, thus keeping the memory constant for any value of $\nstep$. 
Training of the SNFs was performed only for architectures with $\nstep=16,32,64$ for all volumes; the length of the training was 800 epochs for $\nstep=16,32$ and 400 epochs for $\nstep=64$. Results for the smearing parameters on the largest lattice are presented in fig.~\ref{fig:rho}. 

We observed that the values of $\rho$ roughly collapse on the same curve if rescaled with $\nstep$: following this intuition we then implemented a global interpolation of these results obtaining a function valid for all $\nstep$. Possibly due to the relative simplicity of the layers used in this work, we observed no differences when sampling at larger $\nstep$ between a direct training procedure for that architecture and the interpolation carried out uniquely with $\nstep=16,32,64$. In the following all results have been obtained with the latter.
While we performed this training procedure for each volume separately, we noticed that transfer learning between the volumes seems to be possible as well. Namely, we trained $\rho$ on $L/a=12,16,20$ lattices and then applied the results on a SNF with $\nstep=1024$ for the largest lattice $L/a=20$, noticing no discernible differences in the KL divergence or in the $\ESS$. 

\begin{figure}[h]
 \centering
 \begin{subfigure}{.5\textwidth}
 \includegraphics[scale=0.5,keepaspectratio=true]{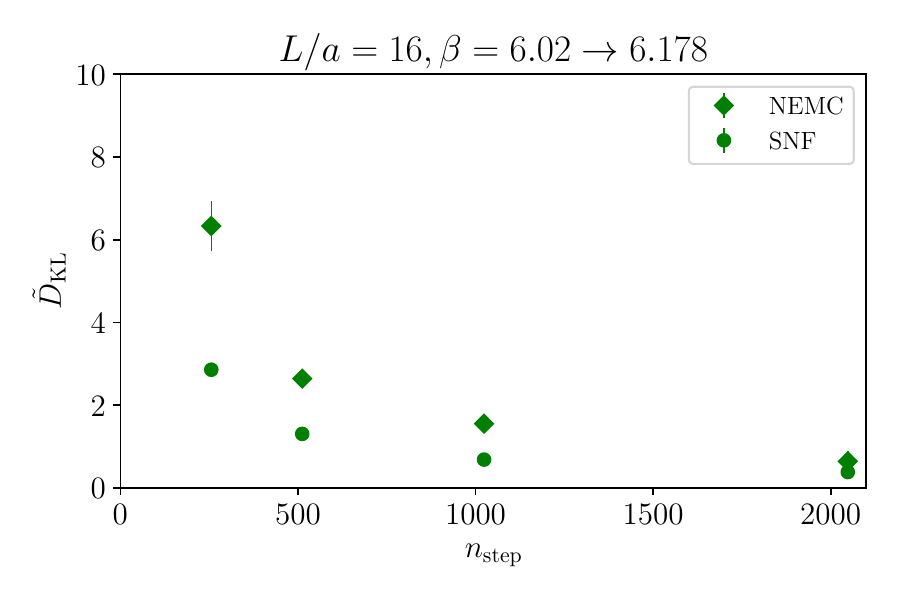}
 \end{subfigure}%
 \begin{subfigure}{.5\textwidth}
 \includegraphics[scale=0.5,keepaspectratio=true]{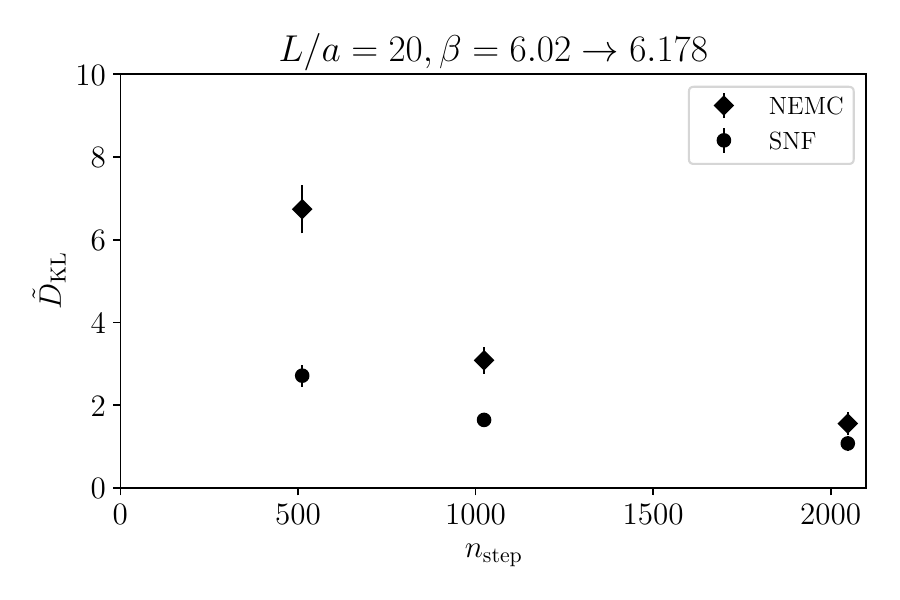}
 \end{subfigure}
 \caption{Results for the KL divergence for $L/a=16$ (left panel) and $L/a=20$ (right panel).}
 \label{fig:dkl_vol}
\end{figure}

In fig.~\ref{fig:dkl_vol} results for the KL divergence of eq.~\eqref{eq:kl} are presented for the two largest volumes, both for the NE-MCMC and for the trained SNFs, for several values of $\nstep$.
For increasing $\nstep$, i.e. for slower evolutions closer to equilibrium, $\DKL$ decreases rather quickly for all architectures and as expected the results indicate a convergence to $0$, i.e. to a reversible transformation. 
Crucially, SNFs are roughly a factor 2 more efficient than the standard NE-MCMC, as the same value of $\DKL$ is reached for about half the value of $\nstep$. 
We stress the fact that this improvement came at a very modest cost: the values of $\rho$ were trained on values of $\nstep$ which are much smaller than the ones tested here. Thus, the computational cost of the training is much smaller than the one used for sampling.

\begin{figure}[h]
 \centering
 \begin{subfigure}{.5\textwidth}
 \includegraphics[scale=0.5,keepaspectratio=true]{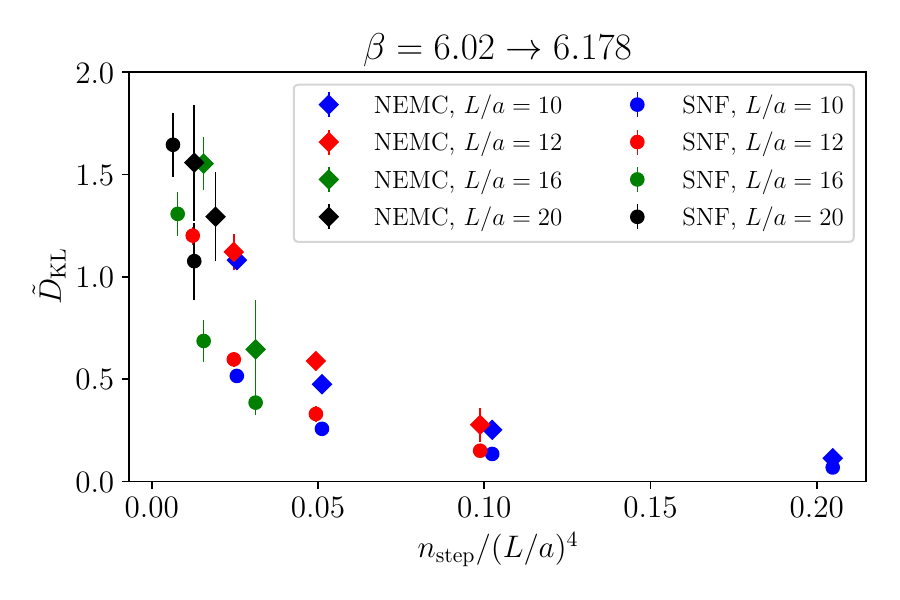}
 \end{subfigure}%
 \begin{subfigure}{.5\textwidth}
 \includegraphics[scale=0.5,keepaspectratio=true]{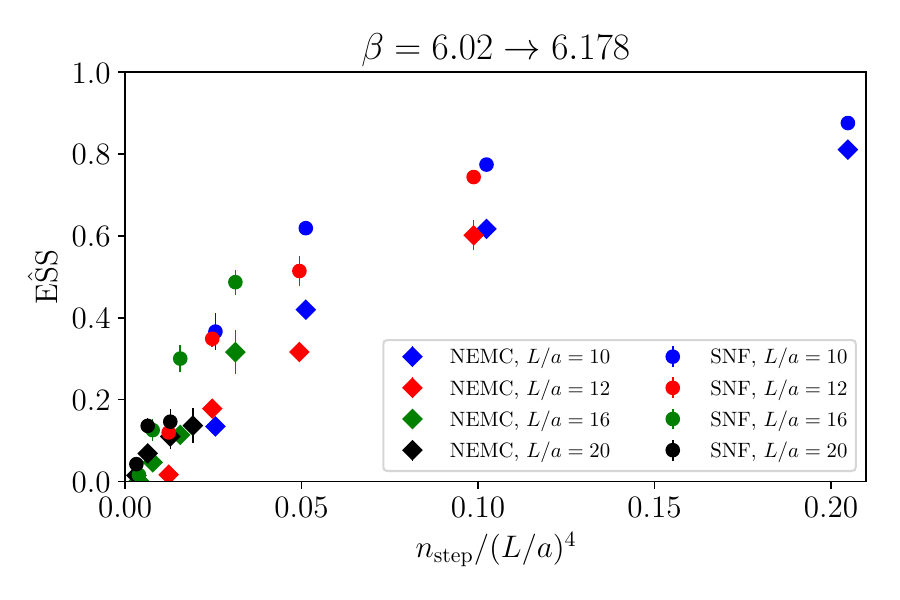}
 \end{subfigure}%
 \caption{Results for the KL divergence (left panel) and for the ESS (right panel) for all the lattice sizes analysed in this study.}
 \label{fig:scaling}
\end{figure}

In fig.~\ref{fig:scaling} we show results for the performance of both for NE-MCMC and for SNF samplers for multiple volumes. 
As already observed in ref.~\cite{Bonanno:2024udh}, if we keep the KL divergence or the $\ESS$ fixed the number of steps $\nstep$ for NE-MCMC scales with the degrees of freedom that are changed throughout a non-equilibrium evolution. In this case $\beta$ is changed over all links, so the d.o.f. of interest go with the volume of the lattice. This is confirmed by the collapse of the points on essentially the same curve for all volumes of interest.
The same scaling appears to hold also for trained SNFs, and this represents one of the main results of this work. All volumes present the same factor of improvement over the NE-MCMC, which is roughly 2 both for the $\DKL$ and the $\ESS$: this suggests that the scaling with the volume (or, more precisely, with the degrees of freedom of the system interested by the evolution) of SNFs is inherited fully from the NE-MCMC.
In this view, the coupling layers simply enhance the non-equilibrium evolution making it more efficient for a limited cost in training.

\section{Conclusions and future work}

In this contribution we have presented the first implementation of Stochastic Normalizing Flows for $\SU(3)$ Yang-Mills theory simulations on the lattice. We tested this approach for a mapping between two probability distributions at different values of $\beta$ for various volumes and we compared it with its purely stochastic counterpart, i.e. non-equilibrium Monte Carlo simulations.

We provided evidence for several promising features of this architecture: first, the training procedure is simple and computationally cheap. The parameters of the coupling layers, through an interpolation, can be rather effortlessly transferred to architectures with larger values of $\nstep$; furthermore, training length is constant for all lattice sizes and transfer of parameters to larger volumes is feasible with no retraining. 
Second, SNFs outperform NE-MCMC by roughly a factor 2 and, even more importantly, appear to inherit the same scaling with the volume of non-equilibrium simulations. Indeed, our results strongly support the idea that larger volumes can be efficiently sampled by scaling $\nstep$ with the number of degrees of freedom interested by the transformation: in this case $\nstep \sim (L/a)^4$ for fixed $\ESS$ or $\DKL$.

The numerical evidence provided in this work represents a crucial step forward in the construction of a flow-based approach that can tackle critical slowing down in lattice gauge theories in four dimensions, with a particular focus towards an efficient mitigation of topological freezing. However, the strategy followed in this work to change $\beta$ over all the links of the lattice becomes very expensive at finer lattice spacings. Indeed, a realistic volume of $L \sim 1.4$ fm at $\beta > 6.4$ would require lattices with $L/a > 30$ and the corresponding $\nstep$ would be in the order of tens of thousands, depending on the starting prior distribution. 
However, the SNF presented in this work can be repurposed for a different approach, i.e. for out-of-equilibrium simulations that change the boundary condition of the system, see refs.~\cite{Bonanno:2024udh, pos_lattice24}: in this mapping $\nstep$ scales only with the size of the boundary changed along the evolution and not with the lattice volume, while exploiting open boundary conditions for the unfreezing of the topological charge.

We conclude by stressing the fact that the current SNF architecture can be systematically improved in several directions. For example, in line with the features of the link-level flow of ref.~\cite{Abbott:2023thq} the stout smearing parameter $\rho$ can be essentially modelled as the output of a simple neural network: the increase in expressivity of the gauge-equivariant coupling layers and its effect on the performances of SNFs is a direction we intend to pursue in future works. 
Moreover, the linear protocol used in this work for both NE-MCMC and SNFs was chosen uniquely for its simplicity. However, in the last few years a large effort has been focused on optimal protocols in non-equilibrium simulations, see ref.~\cite{Blaber_2023} for a review: thus, the development of more efficient protocols in lattice field theory will be a key factor for cost-effective non-equilibrium simulations in the future.

\acknowledgments
We thank S.~Bacchio, C.~Bonanno, M.~Caselle, L.~Giusti, P.~Kessel, S.~Nakajima, K.~Nicoli, M.~Panero, D.~Panfalone, A.~Patella, S.~Schaefer, A.~Tomiya, D.~Vadacchino, L.~Vaitl and L.~Verzichelli for insightful comments and discussions. The numerical simulations were run on machines of the Consorzio Interuniversitario per il Calcolo Automatico dell'Italia Nord Orientale (CINECA). A.~Nada acknowledges support and A.~Bulgarelli and E.~Cellini acknowledge partial support by the Simons Foundation grant 994300 (Simons Collaboration on Confinement and QCD Strings).  A.~Nada acknowledges support from the European Union - Next Generation EU, Mission 4 Component 1, CUP D53D23002970006, under the Italian PRIN “Progetti di Ricerca di Rilevante Interesse Nazionale – Bando 2022” prot. 2022TJFCYB. All the authors acknowledge support from the SFT Scientific Initiative of INFN. 

\bibliographystyle{JHEP}
\bibliography{biblio}

\end{document}